\documentclass[fleqn]{lmcs}
\pdfoutput=1
\usepackage[utf8]{inputenc}

\usepackage{lastpage}
\lmcsdoi{22}{1}{5}
\lmcsheading{}{\pageref{LastPage}}{}{}%
{Jun.~03,~2025}{Jan.~09,~2026}{}

\usepackage[british]{babel}
\usepackage{graphicx,amsfonts,amssymb,microtype,mathdefs,array,mathrsfs,amsthm,relsize}
\usepackage[section]{thmdefs2}
\usepackage[hidelinks]{hyperref}

\let\sc\scshape

\makeatletter
\newcommand\m@thsm@ller[2]{\mbox{\relscale{0.91}$\m@th#1#2$}}
\let\smaller\undefined
\DeclareRobustCommand\smaller[1]{\relax\ifmmode{\mathpalette\m@thsm@ller{#1}}\else{\relscale{0.91}#1}\fi}
\makeatother

\newlist{enuma}{enumerate}{10}
\setlist[enuma]{label={\normalfont(\alph*)}}
\newlist{enumr}{enumerate}{10}
\setlist[enumr]{label={\normalfont(\roman*)}}
\newlist{enumn}{enumerate}{10}
\setlist[enumn]{label={\normalfont(\arabic*)}}

\DeclareRobustCommand*{\dom}{\qopname\relax o{dom}}
\DeclareRobustCommand*{\rng}{\qopname\relax o{rng}}

\newcommand*{\id}{\mathrm{id}}
\newcommand*{\suc}{\mathrm{suc}}

\newcommand*{\fin}{\mathrm{fin}}
\newcommand*{\len}{\mathrm{len}}
\newcommand*{\lex}{\mathrm{lex}}
\newcommand*{\llex}{\mathrm{llex}}
\newcommand*{\pf}{\mathrm{pf}}
\newcommand*{\cn}{\mathrm{cn}}
\newcommand*{\op}{\mathrm{op}}
\newcommand*{\VD}{\smaller{\mathrm{VD}}}
\newcommand*{\FC}{\smaller{\mathrm{FC}}}

\newcommand*{\FOC}{\smaller{\mathrm{FOC}}}
\newcommand*{\MSO}{\smaller{\mathrm{MSO}}}

\newcommand*{\FO}{\smaller{\mathrm{FO}}}

\newcommand*{\?}{\kern .08em}

\let\upsilon\nu

\newcommand\upqed{\vskip-\baselineskip\vskip-\belowdisplayskip}

\keywords{automatic structures, groups, equivalence structures.}

\begin{document}
\title{Simple Classes of Automatic Structures}
\author{Achim Blumensath}
\address{Masaryk University Brno}
\email{blumens@fi.muni.cz}

\begin{abstract}\noindent
We study two subclasses of the class of automatic structures\?: automatic structures
of polynomial growth and Presburger structures. We present algebraic characterisations
of the groups and the equivalence structures in these two classes.
\end{abstract}

\maketitle

\section{Introduction}   

Automatic structures, introduced in~\cite{Hodgson76,Hodgson82,KhoussainovNerode95},
form a quite large class of infinite structures with good algorithmic properties.
Unfortunately these structures are less well-behaved from an algebraic perspective.
In particular, the class lacks good closure properties.
A~persistent and non-trivial problem has been to obtain algebraic characterisations
of when a structure of a certain kind admits an automatic presentation,
or to prove that such a characterisation is not possible.
There is quite a long list of papers devoted to this topic~\cite{Blumensath99,Delhomme04,
KhoussainovRubinStephan05,KhoussainovNiesRubinStephan07,NiesThomas08,Tsankov11,
FinkelTodorcevic13,HuschenbettKartzowLiuLohrey13,AbuZaidGrKrPa14,Rubin21}.
As a further indication of the difficulty of the problem, let us also mention that
it is $\Sigma^1_1$-complete to decide whether a given computable structure is
automatic~\cite{BazhenovHKMN19}.

Because of the difficulty of the full problem, several authors have introduced
subclasses of automatic structures that are simpler to deal with.
The most well-known such class is that of \emph{unary automatic}
structures~\cite{Blumensath99,KhoussainovRubin01}
which, unfortunately, turned out to be too simple to be of much interest.
A second subclass is that of automatic structures of \emph{polynomial growth,}
which was introduced in~\cite{Barany07}, see also~\cite{Huschenbett16,GanardiKhoussainov20}.

In the present article we consider two subclasses of automatic structures
and we give characterisations of two kinds of algebras in these classes\?:
groups and equivalence structures.
The first class is that of structures of polynomial growth.
Automatic equivalence structures of polynomial growth have already been characterised
in~\cite{GanardiKhoussainov20}, but the proof in that article contains an error
caused by a confusion about what kind of coefficients the polynomials under consideration have.
Below we provide a new proof of this result, together with a characterisation of
automatic groups of polynomial growth.

The second class of automatic structures seems to be new\?: structures interpretable
in Presburger arithmetic. This class properly contains the class of structures
of polynomial growth. Also for this class we present characterisations of which groups
and equivalence structures it contains.

The overview of the article is as follows.
We start in Section~\ref{Sect: prelim} with recalling some preliminaries.
Sections \ref{Sect: poly}~and~\ref{Sect: Presburger} introduce the two classes we will be
studying\?: automatic structures of polynomial growth and Presburger structures.
We present several algebraic characterisations of such structures in Sections
\ref{Sect: growth}~and~\ref{Sect: eq.str.}. In the former, we study linear orders and groups,
in the latter, equivalence structures.

\section{Preliminaries}   
\label{Sect: prelim}

Let us fix notation and terminology.
For $k < \omega$, we write $[k] := \{0,\dots,k-1\}$.
We denote the \emph{disjoint union} of two sets $A$~and~$B$ by
$A + B := \{0\} \times A \cup \{1\} \times B$.
The \emph{range} of a function $f : A \to B$ is $\rng f := f[A]$.
We use three different orderings on~$\Sigma^*$\?:
$\leq_\lex$~is the \emph{lexicographic ordering,}
$\leq_\llex$~the \emph{length-lexicographic} one, and
$\leq_\pf$~is the \emph{prefix ordering.}
For further details and all omitted proofs, we refer the reader to~\cite{BlumensathLN1}.

An automatic structure is a relational structure where the universe and each relation
is given by a regular language over some alphabet.
To formally define what it means for a relation to be regular, we encode tuples of words
by a single word over the product alphabet.
\begin{Def}
Let $\Sigma$~be an alphabet and $\Box \notin \Sigma$ a new letter.

(a)
The \emph{convolution} of words $w_0,\dots,w_{n-1} \in \Sigma^*$ is the word
\begin{align*}
  w_0 \otimes\cdots\otimes w_{n-1}
  := 
    \left[\begin{smallmatrix}
      a_{0,0}\\a_{1,0}\\\vdots\\a_{n-1,0}
    \end{smallmatrix}\right]
    \left[\begin{smallmatrix}
      a_{0,1}\\a_{1,1}\\\vdots\\a_{n-1,1}
    \end{smallmatrix}\right]
    \cdots
    \left[\begin{smallmatrix}
      a_{0,l-1}\\a_{1,l-1}\\\vdots\\a_{n-1,l-1}
    \end{smallmatrix}\right]
\end{align*}
over the alphabet $(\Sigma + \{\Box\})^n$ where
\begin{align*}
  l := \max_{i<n} {\abs{w_i}}
\end{align*}
and
$a_{i,j}$~is the $j$-th letter of~$w_i$ or $a_{i,j} := \Box$ if $w_i$~has less than~$j$ letters.

(b)
A relation $R \subseteq \Sigma^* \times\cdots\times \Sigma^*$ is \emph{regular} if
the language
\begin{align*}
  L_R := \set{ w_0 \otimes\cdots\otimes w_{n-1} }{ \langle w_0,\dots,w_{n-1}\rangle \in R }
\end{align*}
is regular.

(c) A relational $\Gamma$-structure $\frakA = \langle A,\bar R\rangle$ is \emph{automatic}
if $\frakA \cong \langle L_D,(L_R)_{R \in \Gamma}\rangle$ where
$L_D$~is a regular language over some alphabet~$\Sigma$ and all relations~$R$
are regular. In this case we call the structure $\langle L_D,(L_R)_{R \in \Gamma}\rangle$
an \emph{automatic presentation} of~$\frakA$.
Structures~$\frakA$ with functions are called automatic,
if the corresponding relational structure
is automatic that is obtained from~$\frakA$ by replacing each function by its graph.
Usually we identify the elements~$a$ of an automatic structure with the words representing them.
The length of this word is denoted by~$\norm{a}$.
\end{Def}

The main reason automatic structures are so well-behaved algorithmically is the fact
that their first-order theory is decidable. In fact, decidability holds for the
following extension of first-order logic.
\begin{Def}
We denote first-order logic by $\FO$, while $\FOC(\sfU)$ is the extension of~$\FO$
by the following quantifiers.
\begin{alignat*}{-1}
  \exists^\infty x&\varphi(x) \quad
    &&\text{`There are infinitely many elements~$x$ satisfying $\varphi$'.} \\
  \exists^{k,m} x&\varphi(x) \quad
    &&\text{`The number of elements~$x$ satisfying $\varphi$ is finite and} \\
    &&&\text{\hphantom{`}congruent $k$ modulo~$m$'.} \\
  \sfU X&\varphi(X) \quad
    &&\text{`There exists an infinite relation~$X$ satisfying $\varphi$'.}
\end{alignat*}
where $x$~is a first-order variable, $X$~a second-order one (not necessarily monadic),
and in the last case we require that $X$~occurs only \emph{negatively} in~$\varphi$.
\end{Def}

Decidability now follows from the following theorem (which is a combination of results
from~\cite{KhoussainovNerode95,Blumensath99,KhoussainovRubinStephan04,KuskeLohrey10}).
\begin{Thm}\label{Thm: FOCU-definable relations are automatic}
Given an automatic structure\/~$\frakA$ (represented by a tuple of automata) and
an\/ $\FOC(\sfU)$-formula~$\varphi(\bar x)$ (without free second-order variables),
one can effectively compute an automaton recognising the relation~$\varphi^\frakA$
defined by~$\varphi$.
\end{Thm}

As a consequence of this result, one can show that automatic structures
are closed under a variety of logical operations. Let us introduce one of them.
\begin{Def}
Let $L$~be a logic like $\FO$ or $\FOC(\sfU)$ and let $k < \omega$.

(a)
A \emph{($k$-dimensional) $L$-interpretation} is defined by a list of formulae
\begin{align*}
  \tau = \langle\delta(\bar x),(\varphi_R(\bar x_0,\dots,\bar x_{n_R-1}))_{R \in \Sigma}\rangle
\end{align*}
where $\Sigma$~is a relational signature,
$\delta$~and~$\varphi_R$ are $L$-formula over some signature~$\Gamma$,
$n_R$~is the arity of the relation~$R$,
and $\bar x,\bar x_i$ are $k$-tuples of variables.

Given a $\Gamma$-structure~$\frakA$, such an interpretation defines a $\Sigma$-structure
\begin{align*}
  \tau(\frakA) := \langle\delta^\frakA,(\varphi_R^\frakA)_{R \in \Sigma}\rangle
\end{align*}
with universe
\begin{align*}
  \delta^\frakA := \set{ \bar a \in A^k }{ \frakA \models \delta(\bar a) }
\end{align*}
and relations
\begin{align*}
  \varphi_R^\frakA := \bigset{ \langle\bar a_0,\dots,\bar a_{n_R-1}\rangle \in A^{kn_R} }
                             { \frakA \models \varphi_R(\bar a_0,\dots,\bar a_{n_R-1}) }\,.
\end{align*}

(b) We say that a structure~$\frakB$ is \emph{$L$-interpretable} in~$\frakA$
if $\frakB = \tau(\frakA)$, for some $L$-interpretation~$\tau$.
\end{Def}

\begin{propC}[\cite{Blumensath99}]\label{Prop: closure under interpretations}
Let $\frakA$~be an automatic structure and $\tau$~an $\FOC(\sfU)$-interpretation.
Then $\tau(\frakA)$ is automatic.
\end{propC}

We can characterise automatic structures via interpretations in various structures.
The most natural ones of these are the following ones.
$\PSet_\fin\langle\omega,{\leq}\rangle$ denotes the structure whose elements
are all finite subsets of~$\omega$ and that has two relations\?:
inclusion~$\subseteq$ and the relation~$\leq$ defined by
\begin{align*}
  A \leq B \quad\defiff\quad
  A = \{a\} \text{ and } B = \{b\} \text{ for some } a \leq b\,.
\end{align*}
The $p$-ary tree $\bigl\langle [p]^*,{\leq_\pf},(\suc_k)_{k < p},{=_\len}\bigr\rangle$
has the prefix-order
\begin{align*}
  u \leq_\pf v \quad\defiff\quad v = ux\,, \quad\text{for some } x \in [p]^*,
\end{align*}
$p$~successor functions $\suc_k(v) := vk$, and the \emph{equal-length} predicate
\begin{align*}
  u =_\len v \quad\defiff\quad \abs{u} = \abs{v}\,.
\end{align*}
Finally, $\bigl\langle \bbN,{+},{}|_p{}\bigr\rangle$ is the expansion of Presburger
Arithmetic by the divisibility predicate
\begin{align*}
  k \mathrel|_p m \quad\defiff\quad k \text{ is a power of } p \text{ dividing } m\,.
\end{align*}
\begin{thmC}[\cite{Blumensath99,ColcombetLoeding07}]%
\label{Thm: interpretations for automatic structures}
Let\/ $\frakA$~be a structure. The following statements are equivalent.
\begin{enumn}
\item $\frakA$~is automatic.
\item $\frakA$ is\/ $\FO$-interpretable in $\PSet_\fin\langle\omega,{\leq}\rangle$
\item $\frakA$ is\/ $\FO$-interpretable in
  $\bigl\langle [p]^*,{\leq_\pf},(\suc_k)_{k < p},{=_\len}\bigr\rangle$,
  for some $p \geq 2$.
\item $\frakA$ is\/ $\FO$-interpretable in $\bigl\langle \bbN,{+},{}|_p{}\bigr\rangle$,
  for some $p \geq 2$.
\end{enumn}
\end{thmC}

\section{Automatic structures of polynomial growth}   
\label{Sect: poly}

Characterising which structures have an automatic presentation is a very hard problem.
To simplify the task, we introduce several subclasses of automatic structures
where it is easier to prove characterisations. We start with the following one,
which was first introduced in~\cite{Barany07}.
\begin{Def}
A language $L \subseteq \Sigma^*$ has \emph{polynomial growth}
if there exists a polynomial~$p(x)$ such that
\begin{align*}
  \bigabs{\set{ w \in L }{ \abs{w} \leq n }} \leq p(n)\,,
  \quad\text{for all } n < \omega\,.
\end{align*}

Similarly, we say that an automatic structure~$\frakA$ has \emph{polynomial growth}
if there exists a polynomial~$p(x)$ such that
\begin{align*}
  \bigabs{\set{ a \in A }{ \norm{a} \leq n }} \leq p(n)\,,
  \quad\text{for all } n < \omega\,.
\end{align*}
In this case we also say that $\frakA$~is \emph{poly-growth automatic.}
\end{Def}
\begin{Exam}
The infinite grid $\langle\bbZ \times \bbZ, E_0,E_1\rangle$ is poly-growth automatic.
(We can represent a point $\langle i,k\rangle \in \bbZ \times \bbZ$ by the word~$a^ib^k$.)
\end{Exam}

The characterisation of automatic structures via interpretations from
Theorem~\ref{Thm: interpretations for automatic structures} can be transferred
to poly-growth automatic structures as follows.
\begin{Thm}\label{Thm: characterisation of poly-growth automatic structures}
Let\/ $\frakA$~be a structure. The following statements are equivalent.
\begin{enumn}
\item $\frakA$~is an automatic structure of polynomial growth.
\item $\frakA$~has an automatic presentation whose universe is a finite union of languages of
  the form
  \begin{align*}
    u_0v_0^*u_1v_1^*\cdots u_{k-1}v_{k-1}^*u_k\,,
    \quad\text{with } u_0,\dots,u_k,v_0,\dots,v_{k-1} \in \Sigma^*.
  \end{align*}
\item $\frakA$~has an automatic presentation whose universe is a finite union of languages of
  the form
  \begin{align*}
    a_0^{m_0}(b_0^{k_0})^*a_1^{m_1}(b_1^{k_1})^*\cdots
      a_{n-1}^{m_{n-1}}(b_{n-1}^{k_{n-1}})^*a_n^{m_n}
  \end{align*}
  for distinct letters $a_0,\dots,a_n,b_0,\dots,b_{n-1}$
  (with distinct members of the union using disjoint alphabets).
\item $\frakA$~is ($k$-dimensionally)\/ $\FO$-interpretable in
  $\langle\bbN, {\leq}, m \divides {\cdot}\rangle$, for some~$m,k$.
\item $\frakA$~is ($k$-dimensionally)\/ $\FO$-interpretable in\/ $\langle\omega,{\leq}\rangle$,
  for some~$k$.
\end{enumn}
\end{Thm}
\begin{proof}
The equivalence (1)~$\Leftrightarrow$~(4) is Theorem~3.3.6 of~\cite{Barany07}.

(1)~$\Leftrightarrow$~(2) follows since, according to~\cite{SzilardYuZhangShallit92},
every regular language of polynomial growth can be written as a finite union of languages
of the form
\begin{align*}
  u_0v_0^*u_1v_1^*\cdots u_{k-1}v_{k-1}^*u_k\,,
  \quad\text{with } u_0,\dots,u_k,v_0,\dots,v_{k-1} \in \Sigma^*,
\end{align*}

(3)~$\Rightarrow$~(2) is trivial.
For (2)~$\Rightarrow$~(3), let $E$~be the relation of all pairs $\langle w,w'\rangle$ where
\begin{align*}
  w &= u_0v_0^{i_0}u_1v_1^{i_1}\cdots u_{n-1}v_{n-1}^{i_{n-1}}u_n \\[0.5em]
\prefixtext{and}
  w' &= a_0^{m_0}b_0^{k_0i_0}a_1^{m_1}b_1^{k_1i_1}\cdots
          a_{n-1}^{m_{n-1}}b_{n-1}^{k_{n-1}i_{n-1}}a_n^{m_n},
\end{align*}
for $i_0,\dots,i_{n-1} < \omega$ and $m_j := \abs{u_j}$ and $k_j := \abs{v_j}$.
Note that $E$~is regular since we can write its convolution~$L_E$ as
\begin{align*}
  &[u_0 \otimes a_0^{m_0}][v_0 \otimes b_0^{k_0}]^*
  [u_1 \otimes a_1^{m_1}][v_1 \otimes b_1^{k_1}]^*
  \cdots {}\\
  &\qquad\qquad\qquad\qquad\quad
    [u_{n-1} \otimes a_{n-1}^{m_{n-1}}][v_{n-1} \otimes b_{n-1}^{k_{n-1}}]^*
  [u_n \otimes a_n^{m_n}]\,.
\end{align*}
We claim that the image of the presentation of~$\frakA$ under~$E$ is again an
automatic presentation of~$\frakA$. Let $R \subseteq A^n$ be a relation of~$\frakA$
and let $R'$~be its image under~$E$. We have to show that~$R'$ is regular.
Let $\Sigma$~be the alphabet used by the given presentation of~$\frakA$
and let $\Gamma$~be the alphabet such that $E \subseteq \Sigma^* \times \Gamma^*$.
By Theorem~\ref{Thm: interpretations for automatic structures},
there exists an $\FO$-formula~$\varphi(\bar x)$ defining~$R$ in the tree
$\bigl\langle \Sigma^*,{\leq_\pf},(\suc_a)_{a \in \Sigma},{=_\len}\bigr\rangle$.
Modifying~$\varphi$ slightly, we obtain an $\FO$-formula~$\varphi'(\bar x)$ defining~$R$ in
\begin{align*}
  \frakT :=
    \bigl\langle (\Sigma + \Gamma)^*,{\leq_\pf},(\suc_a)_{a \in \Sigma + \Gamma},{=_\len}
      \bigr\rangle\,.
\end{align*}
We can define the image~$R'$ inside~$\frakT$ by the formula
\begin{align*}
  \psi(\bar x) := \exists\bar y\Bigl[\varphi'(\bar y) \land \Land_{i<n} Ey_ix_i\Bigr]\,.
\end{align*}
This implies that $R'$~is regular.

(5)~$\Rightarrow$~(4) is trivial.

(4)~$\Rightarrow$~(5)
We can define a $2$-dimensional $\FO$-interpretation of
$\langle\bbN,{\leq},m \divides {\cdot}\rangle$ in $\langle\omega,{\leq}\rangle$ by encoding
the number $mk+i$ by the pair $\langle k,i\rangle$.
This leads to the formulae
\begin{align*}
  \delta(xx') &:= x' < m\,, \\
  \varphi_\leq(xx',yy') &:= x < y \lor [x = y \land x' \leq y']\,, \\
  \varphi_{m \divides {\cdot}}(xx') &:= x' = 0\,.
\end{align*}
\upqed
\end{proof}

\begin{Lem}\label{Lem: closure under unions and products}
The class of poly-growth automatic structures is closed under finite disjoint
unions and finite direct products.
\end{Lem}
\begin{proof}
If we have $\FO$-interpretations of $\frakA$~and~$\frakB$ in $\langle\omega, {\leq}\rangle$,
we can use them to construct interpretations of $\frakA + \frakB$ and $\frakA \times \frakB$
in~$\langle\omega, {\leq}\rangle$.
\end{proof}

\section{Presburger Structures}   
\label{Sect: Presburger}

Our second class is slightly larger than that of the poly-growth automatic structures.
The definition is as follows.
\begin{Def}
(a) A \emph{Presburger structure} is a structure~$\frakA$
for which there exists a (many-dimensional) $\FO$-interpretation of~$\frakA$
in~$\langle\bbN,{+}\rangle$.

(b) We say that a subset $S \subseteq \bbN^n$ is \emph{Presburger-definable} if
it is $\FO$-definable in~$\langle\bbN,{+}\rangle$.
\end{Def}
\begin{Prop}
Every poly-growth automatic structure is a Presburger structure and
every Presburger structure is automatic.
\end{Prop}
\begin{proof}
The first claim follows from
Theorem~\ref{Thm: characterisation of poly-growth automatic structures}
and the fact that there exists an $\FO$-interpretation of
$\langle\bbN,{\leq},m \divides {\cdot}\rangle$ in $\langle\bbN,{+}\rangle$.
The second claim follows by Theorem~\ref{Thm: interpretations for automatic structures}.
\end{proof}

To better understand Presburger structures, we need some results about which kinds of relations
are definable in~$\langle\bbN,{+}\rangle$.
\begin{Def}
(a) A function $\varphi : \bbN^m \to \bbN^n$ is \emph{affine}
if it is of the form
\begin{align*}
  \varphi(\bar x) := u + \sum_{i<m} v_ix_i\,,
  \quad\text{for some } u,v_0,\dots,v_{m-1} \in \bbN^n.
\end{align*}

(b) A set $S \subseteq \bbN^n$ is \emph{semilinear} if it is of the form
\begin{align*}
  S = \rng \varphi_0 \cup\cdots\cup \rng \varphi_{k-1}\,,
\end{align*}
for suitable affine functions $\varphi_0,\dots,\varphi_{k-1}$.

(c) We call $S$~\emph{simple} if it is of the form
$S = \rng \varphi$ for some affine function $\varphi : \bbN^m \to \bbN^n$ such that
the images $\varphi(e_0),\dots,\varphi(e_{m-1})$ of the unit vectors
$e_0,\dots,e_{m-1}$ are linearly independent in~$\bbQ^n$.
\end{Def}

\begin{Def}
(a) For $k,m,p \in \bbN$ with $p > 1$, we write
\begin{align*}
  k \divides_p m \quad\defiff\quad k = p^n \divides m\,, \quad\text{for some } n \in \bbN\,.
\end{align*}

(b) Two natural numbers $k,l \in \bbN$ are \emph{multiplicatively independent}
if the only integer solution to the equation $k^n = l^m$ is $n = 0 = m$.
\end{Def}

It turns out that the Presburger-definable sets are exactly the semilinear ones.
\begin{Thm}\label{Thm: sets definable in Presburger arithmetic}
Let $S \subseteq \bbN^n$. The following statements are equivalent.
\begin{enumn}
\item $S$~is semilinear.
\item $S$~is $\FO$-definable in $\langle\bbN,{+}\rangle$.
\item $S$~is $\FOC(\sfU)$-definable in $\langle\bbN,{+}\rangle$.
\item There are multiplicatively independent numbers $k,l \geq 2$ such that $S$~is
  $\FO$-definable in both $\langle\bbN,{+},{}|_k{}\rangle$ and $\langle\bbN,{+},{}|_l{}\rangle$.
\item There is some $m < \omega$ such that $S$~is quantifier-free definable in the structure
  $\langle\bbN,{+},{\leq}, m \divides {\cdot},0,1\rangle$.
\end{enumn}
\end{Thm}
\begin{proof}
(1)~$\Leftrightarrow$~(2) is a classical result from~\cite{GinsburgSpanier66}.

(5)~$\Rightarrow$~(2) is trivial.

(2)~$\Rightarrow$~(5) holds since the structure
$\langle\bbN,{+},{\leq}, (m \divides {\cdot})_{m<\omega},0,1\rangle$
admits quantifier elimination (see, e.g.,~\cite{Marker02}).

(2)~$\Rightarrow$~(3) is trivial.

(3)~$\Rightarrow$~(4)
Fix an $\FOC(\sfU)$-definable set $S \subseteq \bbN^n$.
The structures $\frakN_k := \langle\bbN,{+},{}|_k{}\rangle$ and
$\frakN_l := \langle\bbN,{+},{}|_l{}\rangle$ have automatic presentations based on,
respectively, the $k$-ary encoding and the $l$-ary encoding.
By Theorem~\ref{Thm: FOCU-definable relations are automatic}, these presentations can be
expanded to ones of, respectively, $\langle\frakN_k,S\rangle$ and $\langle\frakN_l,S\rangle$.
Finally, it follows by Theorem~\ref{Thm: interpretations for automatic structures}
that $S$~is $\FO$-definable in both $\frakN_k$~and~$\frakN_l$.

(4)~$\Rightarrow$~(2) is a classical result by Cobham and Semenov
(see~\cite{DurandRigo21} for an introduction).
\end{proof}

The following result will be useful below to simplify semilinear sets.
\begin{Prop}[Ito~\cite{Ito69}, Eilenberg, Sch\"utzenberger~\cite{EilenbergSchutzenberger69}]%
\label{Prop: making a semilinear set simple}%
Every semilinear set $S \subseteq \bbN^n$ can be written as a disjoint union
of finitely many simple semilinear sets.
\end{Prop}

Our next aim is to derive a bound on the out-degree of a semilinear relation.
\begin{Def}
(a)
A \emph{vector partition function} is a function $f : \bbN^n \to \bbN$ that,
for some matrix $A \in \bbN^{n \times m}$, maps a tuple $\bar x \in \bbN^n$
to the number of tuples $\bar y \in \bbN^m$ such that $\bar x = A\bar y$.
We denote the vector partition function associated with~$A$ by $\psi_A : \bbN^n \to \bbN$.
(Note that not every matrix has an associated vector partition function since the equation
$\bar x = A\bar y$ might have infinitely many solutions.)

(b) A \emph{generalised vector partition function} is a function of the form
\begin{align*}
  g(\bar x) := \sum_{i<s} \psi_{A_i}(\bar x + \bar c_i)\,,
  \quad\text{for } A_i \in \bbN^{n \times m_i} \text{ and } \bar c_i \in \bbN^n.
\end{align*}

(c)
A function $g : \bbN^n \to \bbN$ is a \emph{piecewise polynomial}
if there exists a partition~$\calS$ of~$\bbN^n$ into semilinear sets such that,
for every $S \in \calS$, the restriction $g \restriction S$ is a polynomial
in $\bbQ[x_0,\dots,x_{n-1}]$.
\end{Def}
\begin{Prop}[Sturmfels~\cite{Sturmfels95}]\label{Prop: partition function}
Every vector partition function is piecewise polynomial.
\end{Prop}
\begin{Cor}
Every generalised vector partition function is piecewise polynomial.
\end{Cor}

\begin{Def}
For $\bar x,\bar a \in \bbN^n$, we write
\begin{align*}
  \bar x^{\bar a} := x_0^{a_0}\cdots x_{n-1}^{a_{n-1}}\,.
\end{align*}
\upqed
\end{Def}
\begin{Prop}[Woods~\cite{Woods15}]\label{Prop: out-degree is piecewise quasi-polynomial}
Let $R \subseteq \bbN^k \times \bbN^l$ be a semilinear relation of finite out-degree.
The function $d : \bbN^k \to \bbN$ mapping each tuple $\bar u \in \bbN^k$ to its
$R$-out-degree is a generalised vector partition function.
\end{Prop}
\begin{proof}
We give a simplified proof of the original, stronger statement from~\cite{Woods15}.
With each relation $S \subseteq \bbN^{k+l}$ we associate the formal power-series
\begin{align*}
  f_S(\bar x,\bar y) := \sum_{\langle\bar c,\bar d\rangle \in S} \bar x^{\bar c}\bar y^{\bar d}.
\end{align*}

We can use Proposition~\ref{Prop: making a semilinear set simple} to write
$R = S_0 \cup\dots\cup S_{n-1}$ as a finite disjoint union of simple semilinear sets
$S_i = \rng \varphi_i$. Suppose that
\begin{align*}
  \varphi_i(0) = \bar u_i\bar u'_i
  \qtextq{and}
  \varphi_i(e_j) = v_{i,j}\bar v'_{i,j}\,,
  \quad\text{for } j < s_i\,.
\end{align*}
A direct calculation shows that
\begin{align*}
  f_{S_i}(\bar x,\bar y) =
    \frac{\bar x^{\bar u_i}\bar y^{\bar u'_i}}
         {(1 - \bar x^{\bar v^{\mathstrut}_{i,0}}\bar y^{\bar v'_{i,0}})\cdots
          (1 - \bar x^{\bar v^{\mathstrut}_{i,s_i-1}}\bar y^{\bar v'_{i,s_i-1}})}\,.
\end{align*}
Hence, we obtain
\begin{align*}
  f_R(\bar x,\bar y) =
    \sum_{i < n}
      \frac{\bar x^{\bar u_i}\bar y^{\bar u'_i}}
           {(1 - \bar x^{\bar v^{\mathstrut}_{i,0}}\bar y^{\bar v'_{i,0}})\cdots
            (1 - \bar x^{\bar v^{\mathstrut}_{i,s_i-1}}\bar y^{\bar v'_{i,s_i-1}})}\,,
\end{align*}
which implies that
\begin{align*}
  \sum_{\bar c \in \bbN^k} d(\bar c)\bar x^{\bar c}
  &= \sum_{\bar c \in \bbN^k}
       \bigabs{\set{ \bar d }{ \langle\bar c,\bar d\rangle \in R }}\cdot x^{\bar c} \\
  &= \sum_{\langle\bar c,\bar d\rangle \in R} \bar x^{\bar c} \displaybreak[0]\\
  &= \sum_{\langle\bar c,\bar d\rangle \in R} \bar x^{\bar c}\bar 1^{\bar d} \\
  &= f_R(\bar x,1\dots1) \\
  &= \sum_{i < n}
       \frac{\bar x^{\bar u_i}}
            {(1 - \bar x^{\bar v_{i,0}})\cdots(1 - \bar x^{\bar v_{i,s_i-1}})}\,.
\end{align*}
Since generalised vector partition functions are closed under addition,
it is therefore sufficient to prove that, given a power-series of the form,
\begin{align*}
  \sum_{\bar b} g(\bar b)\bar x^{\bar b}
    = \frac{\bar x^{\bar c}}{(1 - \bar x^{\bar a_0})\cdots (1 - \bar x^{\bar a_{m-1}})}\,,
\end{align*}
the coefficient function $g$~is a generalised vector partition function.
In this case, we obtain
\begin{align*}
  \sum_{\bar b} g(\bar b+\bar c)\bar x^{\bar b}
    &= \frac{1}{(1 - \bar x^{\bar a_0})\cdots (1 - \bar x^{\bar a_{m-1}})} \\
    &= \Bigl[\sum_{\mu_0<\omega} \bar x^{\mu_0\bar a_0}\Bigr]\cdots
       \Bigl[\sum_{\mu_{m-1}<\omega} \bar x^{\mu_{m-1}\bar a_{m-1}}\Bigr] \\
    &= \sum_{\mu_0,\dots,\mu_{m-1}<\omega}
         \bar x^{\mu_0\bar a_0 +\cdots+ \mu_{m-1}\bar a_{m-1}}\,,
\end{align*}
which implies that $g(\bar z+\bar c)$ is equal to the number of tuples $\bar\mu \in \bbN^m$
satisfying
\begin{align*}
  \bar z = \mu_0\bar a_0 +\cdots+ \mu_{m-1}\bar a_{m-1}\,.
\end{align*}
Hence, $g(\bar z) = \psi_A(\bar z+\bar c)$, for some $A$~and~$c$.
\end{proof}
\begin{Exam}
Let $R \subseteq \bbN^2 \times \bbN$ be the set of all triples $\langle a,b,c\rangle$
such that $c$~is an even number with $a \leq c \leq b$. This relation is definable
in~$\langle\bbN,{+}\rangle$ and, hence, semilinear. Its out-degree is
\begin{align*}
  d(a,b) := \begin{cases}
              0                    &\text{if } a > b\,, \\
              \frac{1}{2}(b-a) + 1 &\text{for } a \leq b \text{ and } a,b \text{ even,} \\
              \frac{1}{2}(b-a)     &\text{for } a \leq b \text{ and } a,b \text{ odd,} \\
              \frac{1}{2}(b-a + 1) &\text{otherwise.}
            \end{cases}
\end{align*}
In particular, note that $d \in \bbQ[a,b]$, but $d \notin \bbN[a,b]$.
Finally, note that $d$~is the a vector partition function associated with the equation
\begin{align*}
  \begin{bmatrix}
    a \\ b
  \end{bmatrix}
  =
  \begin{bmatrix}
    0 & 0 & 1 \\ 1 & 2 & 1
  \end{bmatrix}
  \cdot
  \begin{bmatrix}
    x \\ y \\ z
  \end{bmatrix}\,.
\end{align*}
\upqed
\end{Exam}

\section{Growth Arguments}   
\label{Sect: growth}

To prove that certain structures are not poly-growth automatic we can use
the following growth argument, which follows from
Theorem~\ref{Thm: FOCU-definable relations are automatic} together with a pumping argument
originally due to Khoussainov and Nerode~\cite{KhoussainovNerode95}.
\begin{Def}
Let $\frakA$~be a structure and $\varphi(\bar x,y)$ a formula.
For a set $U \subseteq A$ and a number $n < \omega$, we define
the set $N_\varphi(U,n)$ of \emph{reachable elements at distance~$n$} by
\begin{align*}
  N_\varphi(U,0) &:= U\,, \\
\prefixtext{and}
  N_\varphi(U,n+1) &:=
    N_\varphi(U,n) \cup
      \set{ b \in A }
          { \frakA \models \varphi(\bar a,b) \text{ for some } \bar a \subseteq U }\?.
\end{align*}
\upqed
\end{Def}
\begin{Lem}\label{Lem: bounded length increase}
Let\/ $\frakA$~be an automatic structure.
For every\/ $\FOC(\sfU)$-formula~$\varphi(\bar x;\bar z)$ of finite out-degree,
there exists a constant~$k$ such that
\begin{align*}
  \frakA \models \varphi(\bar a;\bar c)
  \qtextq{implies}
  \norm{\bar a} \leq \norm{\bar c} + k\,,
  \quad\text{for all } \bar a,\bar c\,.
\end{align*}
\end{Lem}
\begin{Lem}\label{Lem: growth bound for poly-growth}
Let\/ $\frakA$~be a poly-growth automatic structure, $U \subseteq A$ finite,
and $\varphi$~a formula of finite out-degree. Then there exist constants $d,k > 0$ such that
\begin{align*}
  \abs{N_\varphi(U,n)} \leq n^d+k\,, \quad\text{for all } n < \omega\,.
\end{align*}
\end{Lem}
\begin{proof}
Let $l := \max {\set{ \norm{c} }{ c \in U }}$.
By Lemma~\ref{Lem: bounded length increase}, we can find a constant~$c$ such that
\begin{align*}
  \norm{a} \leq l+cn\,, \quad\text{for all } a \in N_\varphi(U,n)\,.
\end{align*}
By assumption, there exists a polynomial~$p(x)$ such that the universe of~$\frakA$
contains at most~$p(n)$ words of length at most~$n$.
Consequently,
\begin{align*}
  \abs{N_\varphi(U,n)} \leq p(l+cn)\,.
\end{align*}
\upqed
\end{proof}

Our first case study concerns linear orders. We start with ordinals.
\begin{Thm}
An ordinal $\langle\alpha,{\leq}\rangle$ is poly-growth automatic if, and only if,
$\alpha < \omega^\omega$.
\end{Thm}
\begin{proof}
$(\Rightarrow)$ It was shown in~\cite{Delhomme04} that
all automatic ordinals are smaller than~$\omega^\omega$.

$(\Leftarrow)$
It follows by Lemma~\ref{Lem: closure under unions and products}
and Theorem~\ref{Thm: characterisation of poly-growth automatic structures} that the class of
poly-growth automatic ordinals is closed under ordinal addition and multiplication.
Furthermore, $\langle\omega,{\leq}\rangle$ is poly-growth automatic.
\end{proof}

\begin{Def}
Let $\frakA$~be a coloured linear order.

(a) $\frakA$~is \emph{scattered} if the order of the rationals cannot be embedded into~$A$.

(b) $\frakA$~is \emph{regular} if it can be ($1$-dimensionally) $\MSO$-interpreted
in the infinite binary tree $\langle\{0,1\}^*,\suc_0,\suc_1\rangle$.
\end{Def}
\begin{Prop}\label{Prop: scattered regular implies poly-growth automatic}
Let\/ $\frakA$~be a coloured linear order.
If\/ $\frakA$~is regular and scattered, it is poly-growth automatic.
\end{Prop}
\begin{proof}
It is a well-known result (see, e.g., Section~\smaller{VI}.4 of~\cite{BlumensathLN1})
that every scattered regular linear order~$\frakA$ can be constructed from finite linear orders
using finite ordered sums and right multiplication by $\omega$~or~$\omega^\op$.
By Lemma~\ref{Lem: closure under unions and products}
and Theorem~\ref{Thm: characterisation of poly-growth automatic structures},
all of these operations preserve poly-growth automaticity.
\end{proof}
\begin{Exam}
The converse is not true. Let $\langle\omega,{\leq},P\rangle$ be the order with
\begin{align*}
  P := \set{ n(n+1)/2 }{ n < \omega }\,.
\end{align*}
This order has an automatic presentation $\langle a^*b^*,{\leq_\lex},a^*\rangle$
with polynomial growth, but it is not regular.
(It cannot be expressed using the operations from the proof of
Proposition~\ref{Prop: scattered regular implies poly-growth automatic}).
\end{Exam}

Instead of the converse, we can use Lemma~\ref{Lem: growth bound for poly-growth} to prove
the following weaker statement.
\begin{Def}
Let $\calZ$~be the set consisting of all finite linear orders together with
$\omega$,~$\omega^\op$ ($\omega$~with the opposite ordering), and $\bbZ$.
By induction on an ordinal~$\alpha$, we define classes $\mathrm{VD}_\alpha$ of linear orders
as follows.
\begin{align*}
  \VD_0 &:= \{0,1\}\,, \\
  \VD_{\alpha+1} &:=
    \bigset{ \textstyle\sum_{i \in I} \frakA_i }
           { I \in \calZ,\ \frakA_i \in \mathrm{VD}_\alpha }\,, \\
  \VD_\delta &:= \bigcup_{\alpha<\delta} \VD_\alpha\,, \quad\text{for limit ordinals } \delta\,.
\end{align*}
The \emph{$\VD$-rank} $\VD(\frakA)$ of a linear order~$\frakA$ is the least ordinal~$\alpha$
with $\frakA \in \VD_\alpha$.
If no such ordinal exists, we set $\VD(\frakA) := \infty$.
\end{Def}
\begin{Prop}
For every poly-growth automatic linear order\/~$\frakA$, we have $\VD(\frakA) < \omega$.
\end{Prop}
\begin{proof}
For the proof, we introduce a second rank for linear orders.
The \emph{condensation} $\cn(\frakA)$ of a linear oder~$\frakA$ is the quotient~$\frakA/{\sim}$
by the equivalence relation
\begin{align*}
  x \sim y \quad\defiff\quad
  \text{there are only finitely many elements between } x \text{ and } y\,.
\end{align*}
For each ordinal~$\alpha$, we define the $\alpha$-th iteration of~$\cn$ by
\begin{align*}
  \cn^0(\frakA) := \frakA\,,\quad
  \cn^{\alpha+1}(\frakA) = \cn(\cn^\alpha(\frakA))\,,
\end{align*}
and, for a limit ordinal~$\delta$, $\cn^\delta(\frakA)$ is the colimit of the sequence
$(\cn^\alpha(\frakA))_{\alpha<\delta}$.
The \emph{finite condensation rank} $\FC(\frakA)$ of~$\frakA$ is the least ordinal~$\alpha$
such that $\cn^{\alpha+1}(\frakA) = \cn^\alpha(\frakA)$.
It has been shown in~\cite{KhoussainovRubinStephan05} that $\FC(\frakA)$ is finite,
for every automatic linear order~$\frakA$.
Furthermore, it is known that $\FC(\frakA) = \VD(\frakA)$, for every scattered countable
linear order (see, e.g., Section~5.3 of~\cite{Rosenstein82}).

Hence, it is sufficient to show that every poly-growth automatic linear order is scattered.
For a contradiction, suppose that there exists a poly-growth automatic linear order~$\frakA$
that is not scattered. Set $n := \FC(\frakA)$.
Then $\cn^n(\frakA) \cong \langle\bbQ,{\leq}\rangle$.
Since we can $\FOC$-interpret $\cn^n(\frakA)$ in~$\frakA$,
the order $\langle\bbQ,{\leq}\rangle$ is poly-growth automatic.
Let $\varphi(x,y,z)$ be the formula stating that $z$~is the $\leq_\llex$-least element
with $x < z < y$. Then $\varphi$~has finite out-degree and
\begin{align*}
  \abs{N_\varphi(\{a,b\},n)} = 2 + 2^{n-1},
  \quad\text{for } a < b
\end{align*}
(in the structure $\langle\bbQ,{\leq}\rangle$).
A~contradiction to Lemma~\ref{Lem: growth bound for poly-growth}.
\end{proof}

Next, let us take a look at groups and semigroups.
A~characterisation of all finitely generated Presburger groups follows immediately
from the corresponding characterisation of automatic groups.
(Here, an \emph{automatic group} is an automatic structure that happens to be a group.
There is also a commonly used notion of an automatic group due to Thurston~\cite{Epstein92},
which is more restrictive and which we will not be dealing with in this article.)
\begin{Prop}
Let\/ $\frakG$~be a finitely generated group. The following statements are equivalent.
\begin{enumn}
\item $\frakG$~is a Presburger structure.
\item $\frakG$~is automatic.
\item $\frakG$~is virtually abelian.
\end{enumn}
\end{Prop}
\begin{proof}
(2)~$\Leftrightarrow$~(3) has been proved as Theorem~8 of~\cite{OliverThomas05}\?;
(1)~$\Rightarrow$~(2) is trivial\?;
and (3)~$\Rightarrow$~(1) follows from Remark~4 in~\cite{OliverThomas05}
where the authors construct an interpretation of~$\frakG$
in $\langle\bbZ,{+}\rangle$ and, hence, also in $\langle\bbN,{+}\rangle$
(see also Section~\smaller{XII}.9 of~\cite{BlumensathLN1}).
\end{proof}

The class of poly-growth automatic groups turns out to be much smaller.
Before giving the characterisation, let us take a quick look at poly-growth
automatic semigroups.
\begin{Lem}\label{Lem: (N,+) not poly-growth}
Let\/ $\frakS$~be a semigroup such that there exists an embedding of
$\langle\bbN \setminus \{0\},{+}\rangle$ into\/~$\frakS$.
Then\/ $\frakS$~is not poly-growth automatic.
\end{Lem}
\begin{proof}
Suppose that $\frakS = \langle S,{+}\rangle$ is an automatic semigroup into which
$\langle\bbN \setminus \{0\},{+}\rangle$ can be embedded,
and let $c$~be the image of~$1$ under this embedding.
By Lemma~3.2 of~\cite{KhoussainovNiesRubinStephan07}, there exists a constant~$k$ such that
\begin{align*}
  \norm{nc} \leq \norm{c} + k\log_2 n\,,
  \quad\text{for all } n\,.
\end{align*}
It follows that
\begin{align*}
  n \leq 2^{(m - \norm{c})/k}
  \qtextq{implies}
  \norm{nc} \leq m\,.
\end{align*}
Hence, the set $\set{ a \in S }{ \norm{a} \leq m }$ contains at least
$2^{(m - \norm{c})/k}$~elements and $\frakS$~is not of polynomial growth.
\end{proof}

It turns out that the only poly-growth automatic groups are the finite ones.
\begin{Thm}\label{Thm: poly-growth automatic groups}
A group is poly-growth automatic if, and only if, it is finite.
\end{Thm}
\begin{proof}
Let $\frakG = \langle G,{}\cdot{},{}^{-1},e\rangle$ be a poly-growth automatic group.
By Lemma~\ref{Lem: bounded length increase} there exists a constant~$k$ such that
\begin{itemize}
\item for every $a \in G$, there is some $b \in G$ with
  $\norm{a} < \norm{b} \leq \norm{a} + k\,,$
\item $\norm{ab}     \leq \max {\{\norm{a},\norm{b}\}} + k\,,$
\item $\norm{a^{-1}} \leq \norm{a} + k\,.$
\end{itemize}
Setting $m := \norm{e}$, it follows that, for each $n < \omega$,
there exists some element $a_n \in A$ of length
\begin{align*}
  m + 4kn \leq \norm{a_n} < m + 4kn + k\,.
\end{align*}
Set $D_0 := \{e\}$,
\begin{align*}
  C_n &:= \set{ a_0^{s_0}\cdots a_{n-1}^{s_{n-1}} }{ s_0,\dots,s_{n-1} \in \{0,1\} }\,, \\
  D_n &:= \set{ a^{-1}b }{ a,b \in C_n }\,.
\end{align*}
We claim that
\begin{enumr}
\item $\norm{c} < m + 4k(n-1) + 2k$,\quad for all $n > 0$ and $c \in C_n$,
\item $\norm{c} < m + 4kn$,\quad for all $c \in D_n$,
\item $\abs{C_n} = 2^n$.
\end{enumr}
It follows that $G$~contains at least $2^n$ elements of length at most $4kn$.
A~contradiction to the fact that $\frakG$~has polynomial growth.
Hence, it remains to prove the above claims.

\textsc{(i)}
We proceed by induction on~$n$.
For $n = 1$, we have $\norm{e} = m$ and $\norm{a_0} < m+k \leq m+2k$.
For the inductive step, let $c \in C_n$.
If $c \in C_{n-1}$, the claim follows by inductive hypothesis.
Otherwise, we can write $c = da_{n-1}$ with $d \in D_{n-1}$.
Then $\norm{d},\norm{a_{n-1}} < m + 4k(n-1) + k$ implies, by choice of~$k$, that
\begin{align*}
  \norm{c} < m + 4k(n-1) + 2k\,.
\end{align*}

\textsc{(ii)}
Let $a,b \in C_n$. By~\textsc{(i),} we have $\norm{a},\norm{b} < m+4k(n-1)+2k$.
By choice of~$k$, this implies that
\begin{align*}
  \norm{a^{-1}b} < m+4k(n-1)+2k+2k = m+4kn\,.
\end{align*}

\textsc{(iii)}
Suppose that
\begin{align*}
  a_0^{s_0}\cdots a_n^{s_n} = a_0^{t_0}\cdots a_n^{t_n},
  \quad\text{for } s_0,\dots,s_n,t_0,\dots,t_n \in \{0,1\}\,.
\end{align*}
We prove that $s_i = t_i$ by induction on~$n$. Set
\begin{align*}
  b := a_0^{s_0}\cdots a_{n-1}^{s_{n-1}}
  \qtextq{and}
  c := a_0^{t_0}\cdots a_{n-1}^{t_{n-1}}.
\end{align*}
If $s_n = t_n$, we obtain $b = c$ and the claim follows by inductive hypothesis.
Otherwise, we may assume without loss of generality that $s_n = 0$ and $t_n = 1$.
Hence,
\begin{align*}
  b = ca_n
  \qtextq{implies}
  a_n = c^{-1}b \in D_n\,.
\end{align*}
By~\textsc{(iii),} it follows that $\norm{a_n} < m + 4kn$.
A~contradiction to our choice of~$a_n$.
\end{proof}

\section{Equivalence Structures}   
\label{Sect: eq.str.}

The aim of this last section is to prove characterisations both of Presburger equivalence
relations and of poly-growth automatic equivalence relations.
\begin{Def}
(a) An \emph{equivalence structure} is a structure of the
form $\langle A,{\sim}\rangle$ where $\sim$~is an equivalence relation on~$A$.

(b) Given a function $g : \bbN^n \to (\bbN \setminus \{0\}) \cup \{\infty\}$,
we denote by~$\frakE(g)$ the equivalence structure with exactly $\abs{g^{-1}(k)}$ classes
of size~$k$, for each $k \in (\bbN \setminus \{0\}) \cup \{\infty\}$.
\end{Def}

Our aim is to prove the following two characterisations.
\begin{Thm}\label{Thm: Presburger equivalence structures}
An equivalence structure~$\frakA$ is a Presburger structure if, and only if,
\begin{align*}
  \frakA \cong \frakE(g) + \frakC
\end{align*}
where $g$~is a generalised vector partition function and
$\frakC$~is a countable equivalence structure with only infinite classes.
\end{Thm}
\begin{Thm}\label{Thm: characterisation of poly-growth automatic equivalence structures}
An equivalence structure\/~$\frakA$ is poly-growth automatic if, and only if,
it can be written as a finite disjoint union of
\begin{itemize}
\item structures of the form\/~$\frakE(p)$, for polynomials $p \in \bbN[\bar x]$, and
\item countable equivalence structures where every class is infinite.
\end{itemize}
\end{Thm}
\begin{Rem}
Theorem~\ref{Thm: characterisation of poly-growth automatic equivalence structures}
was already stated in~\cite{GanardiKhoussainov20}, but the proof in that article
contained an error\?: for one direction the authors require a polynomial with natural
coefficients, but the other direction only produces polynomials with rational ones.
Below we will present a new, correct proof.
\end{Rem}

We start with a simple lemma that helps us to define interpretations of structures
of the form $\frakE(g)$ in well-ordered structures, i.e., structures where one of
the relations is a well-ordering.
\begin{Lem}\label{Lem: interpreting E(g)}
Let\/ $\frakA$~be a well-ordered structure and $g : \bbN^n \to \bbN$ a function.
There exists an\/ $\FO$-interpretation of\/ $\frakE(g)$ in\/~$\frakA$
if, and only if, there are $k,m < \omega$, an injective function $\sigma : \bbN^n \to A^k$,
and an $\FO$-definable relation $R \subseteq A^k \times A^m$ such that
\begin{align*}
  d_R(\bar a) = \begin{cases}
                  g(\bar c) &\text{if } \bar a = \sigma(\bar c)\,, \\
                  0         &\text{otherwise}\,,
                \end{cases}
\end{align*}
where $d_R$~is the function mapping a tuple~$\bar a$ to its $R$-out-degree.
\end{Lem}
\begin{proof}
$(\Leftarrow)$ Given~$R$, we set
\begin{align*}
  \langle\bar x,\bar y\rangle \sim \langle\bar x',\bar y'\rangle
  \quad\defiff\quad
  \bar x = \bar x'\,.
\end{align*}
Then $\langle R,{\sim}\rangle \cong \frakE(g)$.

$(\Rightarrow)$ Let $\tau = \langle\delta(\bar x),\varphi(\bar x,\bar y)\rangle$
be a $k$-dimensional $\FO$-interpretation of $\frakE(g) = \langle E,{\sim}\rangle$
in~$\frakA$ and let $\upsilon : \delta^\frakA \to E$ be the corresponding isomorphism.
By definition of~$\frakE(g)$, there exists a bijection $\rho : \bbN^n \to E/{\sim}$ such that
\begin{align*}
  \abs{\rho(\bar k)} = g(\bar k)\,, \quad\text{for all } \bar k \in \bbN^n.
\end{align*}
Set ${\approx} := \varphi^\frakA$ and let $P \subseteq \delta^\frakA$ be the set containing
the minimal (w.r.t.\ the lexicographic ordering induced by the well-ordering of~$\frakA$)
element of each $\approx$-class. Then $R := {\approx} \cap (P \times \bbN^m)$ is $\FO$-definable
and the $R$-out-degree of an element $\bar a \in P$ is
\begin{align*}
  \bigabs{[\bar a]_\approx}
  = \bigabs{[\upsilon(\bar a)]_\sim}
  = g\bigl(\rho^{-1}\bigl([\upsilon(\bar a)]_\sim\bigr)\bigr)
  = g\bigl((\rho^{-1} \circ q \circ \upsilon)(\bar a)\bigr)\,,
\end{align*}
where $q : E \to E/{\sim}$ is the projection.
Since the restriction of $\rho^{-1} \circ q \circ \upsilon$ to~$P$ is bijective,
we obtain the desired function~$\sigma$ by setting
\begin{align*}
  \sigma := (\rho^{-1} \circ q \circ \upsilon \restriction P)^{-1} : \bbN^n \to A^k.
\end{align*}
\upqed
\end{proof}

Our characterization of Presburger equivalence structures can now be proved as follows.
\begin{proof}[Proof of Theorem~\ref{Thm: Presburger equivalence structures}]
$(\Leftarrow)$
Note that the equivalence structures $\frakC_1 := \langle \bbN,E_1\rangle$ and
$\frakC_\infty := \langle \bbN^2,E_\infty\rangle$ with
\begin{align*}
  E_1 := \bbN \times \bbN
  \qtextq{and}
  E_\infty := \bigset{ \langle \langle n,i\rangle,\langle n,j\rangle\rangle }{ n,i,j \in \bbN }
\end{align*}
are Presburger structures. ($\frakC_1$~has a single infinite class and
$\frakC_\infty$~has countably infinitely many.)
Since Presburger structures are closed under finite disjoint unions, it therefore remains
to show that $\frakE(g)$~is Presburger, for every generalised vector partition function
\begin{align*}
  g(\bar x) := \sum_{k<s} \psi_{A_k}(\bar x + \bar c_k)\,,
  \quad\text{for } A_k \in \bbN^{n \times m_k} \text{ and } \bar c_k \in \bbN^n.
\end{align*}
Given such a function~$g$, set $m := \max_k m_k$.
The relation
\begin{align*}
  R := \biglset \langle\bar x,\bar y,k\rangle \in \bbN^n \times \bbN^m \times \bbN \bigmset {}
         & k < s,\ A_k\bar y = \bar x + \bar c_k, \\
         & y_i = 0 \text{ for } i \geq m_k \bigrset
\end{align*}
is Presburger definable and the out-degree of $\bar x \in \bbN^n$ is equal to
\begin{align*}
  \sum_{k<s} \psi_{A_k}(\bar x + \bar c_k) = g(\bar x)\,.
\end{align*}
Consequently, we can use Lemma~\ref{Lem: interpreting E(g)} to find an $\FO$-interpretation
of~$\frakE(g)$ in~$\langle\bbN,{+},{\leq}\rangle$.

$(\Rightarrow)$
Suppose that there exists a $k$-dimensional $\FO$-interpretation of~$\frakA$ in
$\langle\bbN,{+}\rangle$. Note that the substructure~$\frakA_0$ of~$\frakA$ consisting
of all finite equivalence classes can be defined by the $\FOC$-formula
\begin{align*}
  \varphi(x) := \neg\exists^\infty y[y \sim x]\,.
\end{align*}
By Theorem~\ref{Thm: sets definable in Presburger arithmetic}, it therefore follows that
$\frakA_0$~is also a Presburger structure. Hence, it is sufficient to prove that
$\frakA_0 \cong \frakE(g)$, for some generalised vector partition function.
Let $P \subseteq A \subseteq \bbN^n$ be the set containing the $\leq_\lex$-minimal element of
every $\sim$-class.
Since $P$~is definable, so is the relation $R := {\sim} \cap (P \times A)$.
It therefore follows by Proposition~\ref{Prop: out-degree is piecewise quasi-polynomial}
that the function $d : \bbN^n \to \bbN$ mapping a tuple~$\bar k$ to its $R$-out-degree is
of the form
\begin{align*}
  p(\bar x) := \sum_{i<s} \psi_{A_i}(\bar x + \bar c_i)\,.
\end{align*}
Since
\begin{align*}
  d(\bar k) = \begin{cases}
                \bigabs{[\bar k]_\sim} &\text{if } \bar k \in P\,, \\
                0                      &\text{otherwise}\,,
              \end{cases}
\end{align*}
we further have $\frakA \cong \frakE(d)$.
\end{proof}

We can make the description in Theorem~\ref{Thm: Presburger equivalence structures}
more explicit by replacing vector partition functions by certain polynomials.
\begin{Def}
A polynomial $p \in \bbQ[x_0,\dots,x_{n-1}]$ is \emph{positive} if the associated
polynomial function $\bbQ^n \to \bbQ$ restricts to a function $\bbN^n \to \bbN \setminus \{0\}$.
\end{Def}
\begin{Lem}\label{Lem: generalised vpf implies union of Z[x]}
Let $g$~be a generalised vector partition function.
Then\/ $\frakE(g)$ can be written as a finite union of structures
of the form~$\frakE(p)$, for positive polynomials~$p$ with integer coefficients.
\end{Lem}
\begin{proof}
By Proposition~\ref{Prop: partition function}, $g \in \bbQ[\bar x]$ is piecewise polynomial.
Hence, there exists a finite partition~$\calS$ of~$\bbN^n$ into semilinear sets and
a family of polynomials $(q_S)_{S \in \calS}$ such that
\begin{align*}
  g \restriction S = q_S\,, \quad\text{for all } S \in \calS\,.
\end{align*}
By Proposition~\ref{Prop: making a semilinear set simple}, we may assume that every $S \in \calS$
is simple. For each $S \in \calS$, fix an injective affine function~$\varphi_S$
with $S = \rng \varphi_S$.
Then $q_S \circ \varphi_S$ is a polynomial in $\bbQ[\bar x]$. Furthermore, we have
\begin{align*}
  \frakE(g) \cong \sum_{S \in \calS} \frakE(q_S \circ \varphi_S)
\end{align*}
by injectivity of~$\varphi_S$.

To conclude the proof, it is therefore sufficient to show that every structure of the form
$\frakE(q)$ with $q \in \bbQ[\bar x]$ can be written as a finite disjoint union of
structures~$\frakE(h)$ with positive $h \in \bbZ[\bar x]$.
We can write $q = \frac{1}{\mu}q_0$ with $q_0 \in \bbZ[\bar x]$ and $0 < \mu < \omega$.
For each tuple $\bar c \in [\mu]^n$, we obtain a polynomial
\begin{align*}
  p_{\bar c}(\bar x) := q(\mu\bar x + \bar c) \in \bbZ[\bar x]\,.
\end{align*}
(Note that the constant term of~$p_{\bar c}$ belongs to~$\bbZ$ since $q$~induces a function
$\bbN^n \to \bbN$.) Furthermore, we have
\begin{align*}
  \frakE(q) \cong \sum_{\bar c \in [\mu]^n} \frakE(p_{\bar c})\,.
\end{align*}
\upqed
\end{proof}

Let us turn to poly-growth automatic equivalence structures. One direction of
Theorem~\ref{Thm: characterisation of poly-growth automatic equivalence structures}
consists of the following lemma.
\begin{Lem}\label{Lem: E(p) poly-growth automatic}
For every positive $p \in \bbQ_{\geq 0}[x_0,\dots,x_{n-1}]$, the structure\/~$\frakE(p)$
is poly-growth automatic.
\end{Lem}
\begin{proof}
Suppose that
\begin{align*}
  p = \tfrac{1}{\mu}\sum_{j<m} \lambda_j\bar x^{\bar a_j}\,,
  \quad\text{for } \mu,\lambda_0,\dots,\lambda_{m-1} \in \bbN\,.
\end{align*}
Let $k_i := \max_j a_{j,i}$~be the maximal exponent of~$x_i$ in~$p$, and
let $R$~be the relation of all tuples
\begin{align*}
    \langle\bar x,\bar y_0\dots\bar y_{n-1},z,w\rangle
       \in \bbN^n \times \bbN^{k_0} \times\cdots\times \bbN^{k_{n-1}} \times \bbN \times \bbN
\end{align*}
such that
\begin{align*}
  z < m\,, \\
  w < \lambda_z\,, \\
  y_{i,j} < x_i\,, &\quad\text{for } i < n \text{ and } j < a_{j,i}\,, \\
  y_{i,j} = 0\,,   &\quad\text{for } i < n \text{ and } a_{j,i} \leq j < k_i\,.
\end{align*}
Then $R$~is definable in $\langle\omega,{\leq}\rangle$ and the $R$-out-degree of
$\bar x \in \bbN^n$ is equal to
\begin{align*}
  \sum_{j<m} \lambda_j\bar x^{\bar a_j}\,.
\end{align*}
Consequently, we can use Lemma~\ref{Lem: interpreting E(g)}
to construct an $\FO$-inter\-pret\-a\-tion of~$\frakE(p_0)$ in $\langle\omega,{\leq}\rangle$,
where $p_0 := \mu p$.

Finally, note that $\frakE(p)$~can be obtained from~$\frakE(p_0)$ by taking every $\mu$-th
element of each equivalence class.
Hence, $\frakE(p)$ is isomorphic to the substructure of~$\frakE(p_0)$ defined by the formula
\begin{align*}
  \varphi(x) := \exists^{0,\mu}y[y \sim x \land y \leq_\llex x]\,.
\end{align*}
We have obtained $\FOC$-interpretations of~$\frakE(p)$ in~$\frakE(p_0)$ and
of~$\frakE(p_0)$ in $\langle\omega,{\leq}\rangle$.
By Theorem~\ref{Thm: characterisation of poly-growth automatic structures}
(and the fact that $\FOC$-interpretations are closed under composition),
it follows that $\frakE(p)$ is poly-growth automatic.
\end{proof}

For the other direction, we need some results about sets $\FO$-definable in the structure
$\langle\omega,{<}\rangle$.
\begin{Def}
(a) For a partial function $f : A \to B$, we denote by $\frakK(f) := \langle A,\ker f\rangle$
the equivalence structure where
\begin{align*}
  \ker f := \set{ \langle a,a'\rangle }{ a,a' \in \dom(f),\ f(a) = f(a') }\,.
\end{align*}

(b)
For a relation $R \subseteq A \times B$, let $\mathrm{fib}_R : B \to \bbN \cup \{\infty\}$
be the function
\begin{align*}
  \mathrm{fib}_R(b) := \bigabs{\set{ a \in A }{ \langle a,b\rangle \in R }}\,.
\end{align*}

(c)
A polynomial~$p(\bar x)$ is \emph{basic} if it can be written
as a sum of products of binomial coefficients of the form
\begin{align*}
  \binom{a_0x_0 +\cdots+ a_{n-1}x_{n-1} + b}{c}
  \quad\text{with } a_0,\dots,a_{n-1},b,c \in \bbN\,.
\end{align*}
\upqed
\vspace*{0.5em}%
\end{Def}

\begin{Rem}
Note that $\frakK(f) \cong \frakE(\mathrm{fib}_f)$.
Hence we can use the former if we want to construct structures of the form~$\frakE(p)$.
\end{Rem}

Similarly to how we can characterise the Presburger-definable relations by the notion
of a semilinear set, we can describe relations definable in $\langle\omega,{\leq}\rangle$
in a purely combinatorial way. We will show below that, for every $n$-ary $\FO$-definable
relation~$R$, there exists some number $s < \omega$ such that we can write~$R$ as
a finite union of equivalence classes of the equivalence relation
\begin{align*}
  \bar a \sim_s \bar b \quad\defiff\quad
  &\text{for all } i < n, \text{ we have } a_i = b_i < s \text{ or } a_i,b_i \geq s\,,
  \text{ and,} \\
  &\text{for all } i,j < n, \text{ one of the following conditions holds\?:} \\
  &{-}\ \ a_i = a_j + k \text{ and } b_i = b_j + k\,, \quad\text{for some } 0 \leq k < s\,, \\
  &{-}\ \ a_i \geq a_j + s \text{ and } b_i \geq b_j + s\,, \\
  &{-}\ \ a_i \leq a_j - s \text{ and } b_i \leq b_j - s\,.
\end{align*}
We call the equivalence classes of this relation \emph{$s$-cells.} Each $s$-cell can uniquely
be described by a permutation $\sigma : [n] \to [n]$ and a function
$d : [n] \to [s] + \{\infty\}$ as follows.
\begin{Def}
Let $s,n < \omega$.
Given a permutation $\sigma : [n] \to [n]$ and a function $d : [n] \to [s] + \{\infty\}$,
we denote by $C(\sigma,d)$ the set of all tuples $\bar a \in \bbN^n$ such that
\begin{alignat*}{-1}
  d(0) &< \infty &&\qtextq{implies} &a_{\sigma(0)} &= d(0)\,, \\
  d(0) &= \infty &&\qtextq{implies} &a_{\sigma(0)} &\geq s\,, \\
  d(i) &< \infty &&\qtextq{implies} &a_{\sigma(i)} &= a_{\sigma(i-1)} + d(i)\,,
  &&\quad\text{for } i > 0\,, \\
  d(i) &= \infty &&\qtextq{implies} &a_{\sigma(i)} &\geq a_{\sigma(i-1)} + s\,,
  &&\quad\text{for } i > 0\,.
\end{alignat*}
Sets of this form are called \emph{$s$-cells.}
\end{Def}
\begin{Exam}
The $4$-cell $C(\id,d)$ associated with the identity permutation and the function
$d : [7] \to [4] + \{\infty\}$ given by
\begin{alignat*}{-1}
  d(0) &:= \infty\,,\quad
  &d(1) &:= 2\,,\quad
  &d(2) &:= 0\,,\quad
  &d(3) &:= \infty\,,\\
  d(4) &:= 1\,,\quad
  &d(5) &:= \infty\,,\quad
  &d(6) &:= 0
\end{alignat*}
contains all tuples $\bar a \in \bbN^7$ satisfying the following inequalities.
\begin{alignat*}{-1}
  a_0 &\geq 4\,,\quad
  &a_2 &= a_1 = a_0 + 2\,,\quad
  &a_3 &\geq a_2 + 4\,,\\
  a_4 &= a_3 + 1\,,\quad
  &a_6 &= a_5 \geq a_4 + 4\,.
\end{alignat*}
\upqed
\end{Exam}

\begin{Lem}\label{Lem: fibre sizes are polynomial}\leavevmode
\begin{enuma}
\item Two $s$-cells are either disjoint or equal.
\item In the structure $\langle\omega,{\leq}\rangle$, every\/ $\FO$-definable relation~$R$
  is a finite union of disjoint $s$-cells, for some $s < \omega$.
\item For every $s$-cell $C(\sigma,d)$, there exists an injective affine function
  $g : \omega^m \to \omega^n$ with $\rng g = C(\sigma,d)$.
\item For all $m,n,s < \omega$, there exist finitely many polynomials
  $p_0,\dots,p_{k-1} \in \bbN[\bar x]$ with the following properties.
  For every $s$-cell $C(\sigma,d) \subseteq \omega^m \times \omega^n$,
  there exists a quantifier-free formula~$\theta(\bar x)$ and some $i \leq k$ such that
  \begin{align*}
    \mathrm{fib}_{C(\sigma,d)}(\bar b) =
      \begin{cases}
        p_i(\bar b) &\text{if } \langle\omega,{\leq}\rangle \models \theta(\bar b)\,, \\
        0           &\text{otherwise}\,,
      \end{cases}
  \end{align*}
  where, in case $i = k$, we use the definition $p_k(\bar x) := \infty$.

  Furthermore, for every affine map~$\varphi$ whose range is included in the set defined
  by~$\theta$, the composition $p_i \circ \varphi$ is a basic polynomial.
\end{enuma}
\end{Lem}
\begin{proof}
(a) Consider two $s$-cells $C(\sigma,d)$ and $C(\sigma',d')$ that share a common
element $\bar a \in C(\sigma,d) \cap C(\sigma',d')$.
Then
\begin{align*}
  a_{\sigma(0)} \leq\cdots\leq a_{\sigma(n-1)}
  \qtextq{and}
  a_{\sigma'(0)} \leq\cdots\leq a_{\sigma'(n-1)}\,,
\end{align*}
which implies that $(a_{\sigma(i)})_{i<n} = (a_{\sigma'(i)})_{i<n}$.
Consequently,
\begin{align*}
  \sigma' = \tau \circ \sigma \qtextq{and} d = d'\,,
\end{align*}
for some permutation~$\tau$ such that $a_{\tau(i)} = a_i$, for all~$i$.
It follows that $C(\sigma,d) = C(\tau \circ \sigma,d) = C(\sigma',d')$.

(b) Since the structure $\langle\omega,{\leq},\suc,0\rangle$ admits quantifier elimination
(see, e.g., Section~3.2 of~\cite{Enderton01}),
the relation~$R$ is a finite union of relations definable by a conjunction of atomic
formulae and their negations. Such a conjunction can be written as a conjunction of
formulae of the form
\begin{align*}
  x_i = x_j + c\,,\quad
  x_i \geq x_j + c\,,\quad
  x_i = c\,,\quad
  x_i \geq c\,,
  \quad\text{for } c \in \bbN\,.
\end{align*}
In particular, it can be written as a finite union of $s$-cells, for some~$s$.
Hence, so can~$R$. Disjointness follows by~(a).

(c) It is sufficient to construct~$g$ for cells of the form $C(\id,d)$
since we then obtain the corresponding function for $C(\sigma,d)$ with an arbitrary
permutation~$\sigma$ by permuting the coordinates of~$g$ in accordance to~$\sigma$.
We construct~$g$ by induction on the dimension~$n$ of~$C(\id,d)$.

If $n = 1$ and $d(0) < s$, we have $C(\id,d) = \{d(0)\}$ and we can set $g : \bbN^0 \to \bbN$
with $g(0) := d(0)$.
If $n = 1$ and $d(0) = \infty$, we have $C(\id,d) = d(0) + \bbN$ and we can set
$g : \bbN \to \bbN$ with $g(x) := x + d(0)$.

For the inductive step, suppose that $n > 1$. By inductive hypothesis, there exists
a function $g' : \bbN^m \to \bbN^{n-1}$ whose range is the projection of $C(\id,d)$
to the first $n-1$ coordinates. Suppose that the components of~$g'$ are
$g_0,\dots,g_{n-2} : \bbN^m \to \bbN$.
We set $g := \langle g_0,\dots,g_{n-2},g_{n-1}\rangle$ where the additional component~$g_{n-1}$
is defined as follows.
If $d(n-1) < s$, we set
\begin{align*}
  g_{n-1}(\bar x) := g_{n-2}(\bar x) + d(n-1)\,.
\end{align*}
If $d(n-1) = \infty$, we set
\begin{align*}
  g_{n-1}(\bar x,y) := g_{n-2}(\bar x) + y + d(n-1)\,.
\end{align*}

(d) It is sufficient to prove the claim for tuples~$\bar b$ with
$b_0 \leq\cdots\leq b_{n-1}$. For other tuples we then obtain the desired polynomials~$p_i$
and formulae~$\theta$ by permuting the variables.
Hence, fix such a tuple~$\bar b$.
We claim that
\begin{align*}
  \mathrm{fib}_{C(\sigma,d)}(\bar b)
  = \binom{b_0 - s_0}{t_0} \cdot
      \prod_{0<j<n} \binom{b_j - b_{j-1} - s_j}{t_j}\,,
\end{align*}
for suitable constants $s_j,t_j < \omega$.
For $j < n-1$, let
\begin{align*}
  I_0     &:= \set{ i < m }{ \sigma(i) \leq \sigma(m+0) }\,, \\
  I_{j+1} &:= \set{ i < m }{ \sigma(m+j) < \sigma(i) \leq \sigma(m+j+1) }\,.
\end{align*}
Then $i \in I_j$ if $b_{j-1} \leq a_{\sigma(i)} \leq b_j$
for some/all tuples~$\bar a$ with $\bar a\bar b \in C(\sigma,d)$.
(To avoid case distinctions, we will use the convention that $b_{-1} := 0$
for the rest of the proof.)
Furthermore, set
\begin{align*}
  I^0_j := \lset i \in I_j \mset {}
               & d(i) = \infty \text{ and there is } i' \in I_j \text{ with } \\
               & i' > i \text{ and } d(i') = \infty \rset\,.
\end{align*}
Note that every~$a_{\sigma(i')}$ with $i' \in I_j \setminus I^0_j$ is at a fixed distance from
some~$a_{\sigma(i)}$ with $i \in I^0_j$.
Hence, to choose a tuple~$\bar a$ with $\bar a\bar b \in C(\sigma,d)$ amounts to choosing
the values for~$a_{\sigma(i)}$ with $i \in I^0_0 \cup\dots\cup I^0_{n-1}$.
There are
\begin{align*}
  t_j := \abs{I^0_j}
\end{align*}
such elements~$a_{\sigma(i)}$ with $i \in I^0_j$, and
the number of choices for each of them is equal to $b_j - b_{j-1} - s_j$ where
\begin{align*}
  s_j := \sum {\set{ d(i) }{ i \in I^0_j,\ d(i) \neq \infty }}
\end{align*}
is the number of choices that are inadmissible because they are too close to some other element.
Consequently, there are
\begin{align*}
  \binom{b_j - b_{j-1} - s_j}{t_j}
\end{align*}
choices for the part of~$\bar a$ between $b_{j-1}$ and $b_j$.

Having established the above claim, it follows from its proof that we can use the formula
\begin{align*}
  \theta(\bar x) :=
    x_0 \geq s_0 &\land \Land_{j<n-1} x_{j+1} \geq x_j + s_j \\
    &\land \Land {\set{ x_j = x_{j-1} + s_j }{ t_j = 0 }}\,.
\end{align*}

Finally, let $\varphi$~be an affine map whose range is included in the set~$S$ defined
by~$\theta$ and suppose that the associated coordinate maps are
\begin{align*}
  \varphi_j(\bar y) = \sum_i a_{ji}y_i + c_j
  \quad\text{with coefficients } a_{ji},c_j \in \bbN\,.
\end{align*}
Since $\rng \varphi \subseteq S$, we have
\begin{align*}
  \varphi_j(\bar y) - \varphi_{j-1}(\bar y) - s_j \geq 0\,,
  \quad\text{for all } \bar y\,.
\end{align*}
Consequently,
\begin{align*}
  \sum_i (a_{j,i} - a_{j-1,i})y_i + (c_j - c_{j-1}) \geq s_j\,,
  \quad\text{for all } \bar y\,,
\end{align*}
which implies that
\begin{align*}
  \alpha_{j,i} &:= a_{j,i} - a_{j-1,i} \geq 0\,, \quad\text{for all } i\,, \\
  \beta_j    &:= c_j - c_{j-1} - s_j \geq 0\,.
\end{align*}
Hence, we have
\begin{align*}
  &\mathrm{fib}_{C(\sigma,d)}(\varphi(\bar y)) \\
  &\quad{}= \binom{\varphi_0(\bar y) - s_0}{t_0} \cdot
      \prod_{0<j<n} \binom{\varphi_j(\bar y) - \varphi_{j-1}(\bar y) - s_j}{t_j} \\
  &\quad{}= \binom{\sum_i \alpha_{0,i} y_i + \beta_0}{t_0} \cdot
      \prod_{0<j<n} \binom{\sum_i \alpha_{j,i} y_i + \beta_j}{t_j}\,,
\end{align*}
which is basic.
\end{proof}

\begin{Lem}\label{Lem: poly-growth automatic eq. structures with finite classes}
Let\/ $\frakA$~be an equivalence structure with no infinite classes.
The following statements are equivalent.
\begin{enumn}
\item $\frakA$ is poly-growth automatic.
\item $\frakA \cong \frakK(f)$ for some partial function $f : \omega^m \to \omega^n$
  that is\/ $\FO$-definable in $\langle\omega,{\leq}\rangle$.
\item $\frakA$ is a finite disjoint union of structures of the form $\frakE(p)$,
  for some polynomial $p \in \bbN[\bar x]$.
\end{enumn}
\end{Lem}
\begin{proof}
(3)~$\Rightarrow$~(1)
We have seen in Lemma~\ref{Lem: E(p) poly-growth automatic} that
every structure~$\frakE(p)$ with $p \in \bbN[\bar x]$ is poly-growth automatic.
Consequently, so is every finite disjoint union of such structures.

(1)~$\Rightarrow$~(2)
Suppose that $\frakA = \langle A,{\sim}\rangle$ is poly-growth automatic
and let $f : A \to A$ be the function mapping each element $a \in A$ to the $\leq_\llex$-minimal
element of its $\sim$-class.
Since $\frakA$~is $\FO$-interpretable in $\langle\omega,{\leq}\rangle$, we can
regard~$f$ as a partial function $\omega^k \to \omega^k$, for some~$k$.
It follows that $\frakA \cong \frakK(f)$.

(2)~$\Rightarrow$~(3)
Let $f : \omega^m \to \omega^n$ be a definable partial function.
Note that
\begin{align*}
  \frakK(f) \cong \frakE(\mathrm{fib}_f)\,.
\end{align*}
For $s,n < \omega$, we denote by~$\calC^s_n$ the set of all $s$-cells of dimension~$n$.
We can use Lemma~\ref{Lem: fibre sizes are polynomial}\,(b) to find some constant~$s$
such that we can write (the graph of)~$f$ as a disjoint union of $s$-cells.

By Lemma~\ref{Lem: fibre sizes are polynomial}\,(d), there exist a finite set~$\calP$
of polynomials and a finite set~$\Theta$ of $\FO$-formulae such that, for every
$C \in \calC^s_{m+n}$, there is some $p \in \calP$ and some $\theta \in \Theta$ such that
\begin{align*}
  \mathrm{fib}_C(\bar b) = p(\bar b)\,,
  \quad\text{for all } \bar b \in \omega^n \text{ satisfying } \theta\,.
\end{align*}
Using Lemma~\ref{Lem: fibre sizes are polynomial}\,(b) again, we obtain a constant~$t$
such that the relations defined by the formulae in~$\Theta$ are unions of $t$-cells.
It follows that there exists a function $\pi_C : \calC^t_m \to \calP$ such that
\begin{align*}
  \mathrm{fib}_C \restriction D = \pi_C(D)\,,
  \quad\text{for all } C \in \calC^s_{m+n} \text{ and } D \in \calC^t_m\,.
\end{align*}
Consequently,
\begin{align*}
  \mathrm{fib}_f \restriction D
  = \sum {\set{ \pi_C(D) }{ C \in \calC^s_{m+n},\ C \subseteq f }}\,,
  \quad\text{for every } D \in \calC^t_m\,,
\end{align*}
which is a polynomial in $\bbQ[\bar x]$.

For each $D \in \calC^t_m$, fix an affine map~$\varphi_D$ with $\rng \varphi_D = D$. Then
\begin{align*}
  \frakE(\mathrm{fib}_f) = \sum_{D \in \calC^t_m} \frakE(\mathrm{fib}_f \circ \varphi_D)
\end{align*}
and it follows by Lemma~\ref{Lem: fibre sizes are polynomial}\,(d) that
each map $g_D := \mathrm{fib}_f \circ \varphi_D$ is a basic polynomial.

Fix a number $c \in \bbN$ such that $c \geq k$, for every binomial coefficient
$\binom{\ldots}{k}$ appearing in~$g_D$ and set
\begin{align*}
  g'_D(x_0,\dots,x_{n-1}) := g_D(x_0 + c,\dots,x_{n-1} + c)\,.
\end{align*}
Then
\begin{align*}
  \frakE(g'_D) \cong \frakE(g_D)
\end{align*}
and every binomial coefficient appearing in~$g'_D$ is of the form
\begin{align*}
  &\binom{a_0x_0 +\cdots+ a_{n-1}x_{n-1} + b}{k} \\
  &\quad{} = \frac{1}{k!}\prod_{i<k} (a_0x_0 +\cdots+ a_{n-1}x_{n-1}+(b-i))
  \quad\text{with } b \geq k\,.
\end{align*}
In particular $g'_D \in \bbQ_{\geq 0}[\bar x]$.

Finally, fix a number~$d$ such that $d\cdot g'_D \in \bbN[\bar x]$ and set
\begin{align*}
  g''_D(x_0,\dots,x_{n-1}) := g'_D(dx_0,\dots,dx_{n-1})\,.
\end{align*}
Then
\begin{align*}
  \frakE(g''_D) \cong \frakE(g'_D)
  \qtextq{and}
  g''_D \in \bbN[\bar x]\,.
\end{align*}
Since
\begin{align*}
  \frakK(f) \cong \frakE(\mathrm{fib}_f)
            \cong \sum_{D \in \calC^t_m} \frakE(\mathrm{fib}_f \circ \varphi_D)
            \cong \sum_{D \in \calC^t_m} \frakE(g''_D)\,,
\end{align*}
the claim follows.
\end{proof}

\begin{proof}[Proof of Theorem~\ref{Thm: characterisation of poly-growth automatic equivalence structures}]
$(\Rightarrow)$ Let $\frakA$~be poly-growth automatic. We decompose it as
$\frakA = \frakB + \frakC$ where $\frakB$~is an equivalence structures with
only finite classes and $\frakC$~is one with only infinite classes.
Since $B$~and~$C$ are $\FOC$-definable, it follows that
$\frakB$~and~$\frakC$ are poly-growth automatic.
Furthermore, we can use Lemma~\ref{Lem: poly-growth automatic eq. structures with finite classes}
to decompose~$\frakB$ into a disjoint union of structures of the form~$\frakE(p)$
with $p \in \bbN[\bar x]$.

$(\Leftarrow)$ We have seen in Lemma~\ref{Lem: E(p) poly-growth automatic} that
every structure of the form $\frakE(p)$ with $p \in \bbN[\bar x]$ is poly-growth automatic.
Furthermore, the equivalence structures $\frakA_1 := \langle 0^*,E_1\rangle$ and
$\frakA_\infty := \langle 0^*1^*,E_\infty\rangle$ with
\begin{align*}
  E_1 := 0^* \times 0^*
  \qtextq{and}
  E_\infty := \bigset{ \langle 0^n1^k,0^n1^l\rangle }{ n,k,l < \omega }
\end{align*}
are poly-growth automatic. ($\frakA_1$~has a single infinite class and
$\frakA_\infty$~has countably infinitely many.) The claim follows
since the class of poly-growth automatic structures is closed under finite disjoint unions.
\end{proof}

\bibliographystyle{alphaurl}
\bibliography{Polynomial}

\end{document}